\documentclass[prb,twocolumn,showpacs,preprintnumbers,amsmath,amssymb]{revtex4}

\usepackage{graphicx}       
\usepackage{dcolumn}        

\newcommand{\beq}{\begin{equation}}
\newcommand{\eeq}{\end{equation}}
\newcommand{\beqa}{\begin{eqnarray}}
\newcommand{\eeqa}{\end{eqnarray}}

\newcommand{\kvec}{{\bf k}}
\newcommand{\G}{{\cal G}}
\newcommand{\qvec}{{\bf q}}
\newcommand{\pvec}{{\bf p}}
\newcommand{\Rvec}{{\bf R}}
\newcommand{\rvec}{{\bf r}}

\begin{document}
\title{
Effect of mesoscopic inhomogeneities on local tunnelling density of states
}

\author{L. Dell'Anna$^{1}$, J. Lorenzana$^{1}$, M. Capone$^{1,2}$, 
C. Castellani$^{1}$,  and M. Grilli$^{1}$}
\affiliation{$^1$Istituto Nazionale di Fisica della Materia,
Unit\`a Roma 1 and SMC Center, and Dipartimento di Fisica\\
Universit\`a di Roma "La Sapienza" piazzale Aldo Moro 5, I-00185 Roma, Italy}
\affiliation{$^2$Enrico Fermi Center, Rome, Italy}

\begin{abstract}
We carry out a theoretical analysis of the momentum dependence of the
 Fourier-transformed
local density of states (LDOS) in the superconducting cuprates
within a model considering the interference of
quasiparticles scattering on quenched impurities. The impurities introduce
an external scattering potential, which is either nearly local in space or it
can acquire a substantial momentum dependence due to a possible 
strong momentum dependence of the electronic screening near 
a charge modulation instability. The key new effect that we introduce is
an additional mesoscopic disorder aiming to reproduce the
inhomogeneities experimentally observed in scanning tunnelling microscopy.
The crucial effect of this mesoscopic disorder is to give rise to point-like
spectroscopic features, to be contrasted with the curve-like shape of
the spectra previously calculated within the interfering-quasiparticle
schemes. It is also found that stripe-like charge modulations
play a relevant role to correctly reproduce all the spectral features of the
experiments.

\end{abstract}
\date{\today}
\maketitle

\section{Introduction}
In the last few years scanning tunneling microscopy (STM) measurements have become a 
most valuable tool in investigating the physical properties of superconducting
cuprates. Although, like angle-resolved photoemission spectroscopy (ARPES),
this technique is mostly sensitive to the surfaces, the layered structure of the
cuprates suggests that it may be representative of the 
bulk properties. In particular, in the recent years Fourier-transformed (FT) STM spectra
showed a rich momentum structure in the local density of states (LDOS) and
raised a strong debate on what elementary excitations produce such structures
\cite{hoffman1,hoffman2,howald1,howald2,mcelroy,yazdani}. To be specific the dot-like patterns
observed in FT-STM spectra have been attributed either to local charge-spin
order with pinned static collective excitations \cite{howald2} or to interference
effects between impurity-scattered quasiparticles (QP) \cite{hoffman1,hoffman2,mcelroy}.
Several theoretical analyses have tried so far to consider these mechanisms and to
relate them to the experimental observations. Grossly speaking, it is found that
pinned collective textures may account for the rather punctual (although obviously
broadened) character of the LDOS patterns in momentum space 
\cite{polkovnikov,zhu,chen,sachdev,halperin}.
However, the predicted patterns show too weak an energy dependence, which contrasts with the
substantial dispersion of most of the experimentally detected spots. On the other hand,  
the theoretical analyses based on QP interference generically produce 
dispersive LDOS patterns, but at a given energy the high intensity regions form extended
curves in $k$ space \cite{wang,zhang,pereg,capriotti}, which hardly
resemble the experimental spot-like intensity patterns. Moreover
the STM measured dispersion curves do not
properly match the ARPES-determined QP dispersions and tend to ``flatten''
at wave-vectors typical of charge/spin order. Finally, the weight of QP peaks in ARPES is weak
and strongly depends on temperature. Therefore the QP interference effects should 
disappear upon approaching $T_c$\cite{capriotti,kivelson}.
This seems to be the case for many of the
structures, but experimentally it is also
found that some of the k-space features persists above $T_c$\cite{yazdani}.
Therefore neither of the two pictures is fully satisfactory and one still needs
to reconcile these analyses with experiments.
In this paper we precisely aim to perform a systematic analysis at low temperature of the QP
picture to determine whether or not physically sensible mechanisms (atomic form factors, disorder,
multiple scattering, mesoscopic inhomogeneities, charge-ordering instabilities) 
can turn the curve-like QP spectra into more spot-like patterns.
We consider the structure factors due to
the short-distance structure of the Wannier orbitals, the second-order impurity scattering
processes, the effect of magnetic impurities and, most importantly, the effect of
mesoscopic inhomogeneities. In this way we succeed in smoothing most of the curve-like
LDOS features into broad peaky structures. However, a detailed comparison of the experimental
figures with our calculated spectra shows that some features cannot be properly reproduced.
A better agreement is instead obtained when additional effects from charge-density modulations
are considered. Therefore, our analysis shows that in the real cuprate systems
there must be a coexistence between dispersive QPs (producing dispersive interference patterns
broadened by  inhomogeneity effects) and incipient static local charge order 
responsible for the enhancement of 
non-dispersive peaks in specific regions of the $k$ space.  

The paper is organized as follows. Section II contains the model and the description of
the approach. In Section III the mesoscopic inhomogeneities are introduced in the model
and their effects are described. In Section IV
 we present our results, while our concluding remarks are contained
in Section V. 

\section{The model and the technique}
\subsection{The model}
In STM experiments, the LDOS is measured with atomic resolution on the points $\rvec$ of 
a field of view $\ell\times\ell$ with $\ell\sim 640$\AA~
(several hundreds
of lattice unit cells). Once the LDOS $N(\rvec,\omega)$ is obtained,
its Fourier transform is a function of the  momenta $(q_x,q_y)=(n_x,n_y)2\pi/\ell$,
yielding the wave-vector power spectrum
\beq
P(\qvec,\omega)\equiv |N(\qvec,\omega)|^2/\ell^2.
\eeq
Notice that, since the $\rvec$
positions of the STM scans are denser than the atomic positions,
the $\qvec$ momenta are not restricted to the first Brillouin zone. 
In the following we work on an infinite lattice defined on atomic positions $\Rvec$ 
(in units of the lattice spacing $a$)
and use a field of view of size $L\times L$, where $L=\ell/a$.
Therefore the momenta are restricted to the first Brillouin zone. This restriction is
relaxed in Appendix A by including the atomic form factors encoding the orbital
subatomic structure. 

To consider the elastic scattering of the quasiparticles on quenched impurities,
we introduce a (weak) external potential 
$\epsilon(\Rvec)=\sum_{i=1}^{N_i} \epsilon(\Rvec-\Rvec_i)$
due to the local potential of $N_i$ impurities randomly located on sites ${\bf R}_i$
of the $L\times L=N$ two-dimensional system. Although most of the expressions
below stay valid for a general form of the impurity potential, to be more specific
we consider the form
\beq 
\epsilon({\bf R})=\sum_i^{N_i} V_0 \delta ({\bf R}-{\bf R}_i)
\eeq
which in momentum space reads
\beq
\epsilon (\qvec)= {V_0}\sum_{i=1}^{N_i} e^{i \qvec \Rvec_i}
\eeq
Aiming to perturbatively calculate the corrections induced by the
impurity potential on the LDOS, we need to calculate the electron
Green's function in the superconducting state. Thus it is convenient to
introduce fermionic Nambu spinors to write the Green's functions in matrix form 
\beq
\nonumber
\hat\G=\left(
\begin{array}{cc}
{\cal{G}}(\kvec,\omega) & {\cal{F}}(\kvec,\omega)\\
{\cal{F}}(\kvec,\omega) & {\cal{G}}(\kvec,-\omega)\\
\end{array}
\right)
\eeq
where the normal and anomalous Green functions are given by 
\begin{equation}
\label{G11}
{\cal{G}}(\kvec,\omega)=  \frac{\omega+\epsilon_\kvec}{\omega^2-E_\kvec^2}\;\;\;\;\;
{\cal{F}}(\kvec,\omega)=\frac{\Delta_\kvec}{\omega^2-E_\kvec^2}.
\end{equation}
Also the scalar (i.e., non magnetic) impurity potential can be put
in matrix form
 \begin{eqnarray*}
\hat{\epsilon}_s(\qvec)=\epsilon(\qvec)\left(
\begin{array}{cc}
1 & \phantom{-}0\\
0 & -1\\
\end{array}
\right).
\end{eqnarray*}
allowing one to define the quantity
\begin{eqnarray}
\label{Lambda}
\Lambda_-(\qvec,\omega)&=& \int \frac{d^2k}{(2\pi)^2} \times \\
&&{\cal{G}}(\kvec,\omega)
{\cal{G}}(\kvec+\qvec,\omega)-{\cal{F}}
(\kvec,\omega){\cal{F}}(\kvec+\qvec,\omega) \nonumber\\
&=&\int \frac{d^2k}{(2\pi)^2} \frac{(\omega+\epsilon_\kvec)(\omega+\epsilon_{\kvec+\qvec})
-\Delta_\kvec \Delta_{\kvec+\qvec}}{(\omega^2-E_\kvec^2)(\omega^2-E_{\kvec+\qvec}^2)}\nonumber
\end{eqnarray}
as the $(1,1)$ element of the matrix product 
\beq
\nonumber
\left(
\begin{array}{cc}
{\cal{G}}(\kvec,\omega) & {\cal{F}}(\kvec,\omega)\\
{\cal{F}}(\kvec,\omega) & {\cal{G}}(\kvec,-\omega)\\
\end{array}
\right)\times 
\eeq
\beq
\left( 
\begin{array}{cc}
1 & \phantom{-}0\\
0 & -1\\
\end{array}
\right)
\left(
\begin{array}{cc}
{\cal{G}}(\kvec+\qvec,\omega) & {\cal{F}}(\kvec+\qvec,\omega)\\
{\cal{F}}(\kvec+\qvec,\omega) & {\cal{G}}(\kvec+\qvec,-\omega)\\
\end{array} 
\right).
\eeq
The integral in Eq.~(\ref{Lambda}) emphasizes the fact that we are
working on an infinite lattice. 

Since the LDOS is obtained from the imaginary part of the
(1,1) element of the Green function matrix, at first order in
the impurity potential,
the correction to the LDOS can be written as
\begin{eqnarray}
\label{sum}
&&N^{(1)}(\Rvec,\omega)=\Im  \int \frac{d^2q}{(2\pi)^2}e^{i {\bf q}\cdot {\bf R}} \, \epsilon({\bf q})
\Lambda_-(\qvec,\omega) \nonumber \\
&=&\int \frac{d^2q}{(2\pi)^2}\\ 
&&\left\{\left(\epsilon^{\Re}(\qvec)\cos(\qvec\cdot \Rvec)+\epsilon^{\Im}(\qvec)
 \sin(\qvec\cdot \Rvec)\right)\,\Im (\Lambda_- (\qvec,\omega))\right.\nonumber\\
&+&\left.\left(\epsilon^{\Re}(\qvec)\sin(\qvec\cdot \Rvec)+\epsilon^{\Im}(\qvec)
 \cos(\qvec\cdot \Rvec)\right)\,\Re (\Lambda_- (\qvec,\omega))\right\}. \nonumber
\end{eqnarray}
Moreover, since
\begin{eqnarray}
\epsilon^{\Re}(\qvec)=\frac{1}{2}\left[\epsilon(\qvec)+\epsilon(\qvec)^*\right]=
\frac{1}{2}\left[\epsilon(\qvec)+\epsilon(-\qvec)\right]\\
\epsilon^{\Im}(\qvec)=\frac{1}{2}\left[\epsilon(\qvec)-\epsilon(\qvec)^*\right]=
\frac{1}{2}\left[\epsilon(\qvec)-\epsilon(-\qvec)\right]
\end{eqnarray}
then $\epsilon^{\Re}(\qvec)$ is the symmetric part of $\epsilon(\qvec)$ with respect to 
$\qvec\rightarrow -\qvec$ while $\epsilon^{\Im}(\qvec)$ is the antisymmetric one. Both 
$\Im (\Lambda_- (\qvec))$ 
and $\Re (\Lambda_- (\qvec))$ are symmetric with respect to the same transformation 
so the only term that survives 
under the sum over $\qvec$ is the first one in (\ref{sum}), then 
\begin{eqnarray}
&&N^{(1)}(\Rvec,\omega)=\int \frac{d^2q}{(2\pi)^2}\\
&&\left(\epsilon^{\Re}(\qvec)\cos(\qvec\cdot \Rvec)
+\epsilon^{\Im}(\qvec) \nonumber
 \sin(\qvec\cdot \Rvec)\right)\,\Im (\Lambda_- (\qvec,\omega))
\end{eqnarray}
If there is inversion symmetry (for example one symmetric  
impurity at $\Rvec =(0,0)$) so that $\epsilon(\Rvec)=\epsilon(-\Rvec)$ then 
$\epsilon^{\Im}(\qvec)=0$ 
and $\epsilon(\qvec)=\epsilon^{\Re}(\qvec)$ and one simply has
\begin{eqnarray}
N^{(1)}(\Rvec,\omega)=\int \frac{d^2q}{(2\pi)^2}\epsilon(\qvec)\cos(\qvec\cdot \Rvec)\,\Im (\Lambda_- 
(\qvec,\omega))
\end{eqnarray}
For $N_i$ delta-like impurities one has
\begin{eqnarray}
\epsilon^{\Re}(\qvec)={V_0}\sum^{N_i}_{i=1} \cos(\qvec\cdot \Rvec_i)\\
\epsilon^{\Im}(\qvec)={V_0}\sum^{N_i}_{i=1} \sin(\qvec\cdot \Rvec_i)
\end{eqnarray}
where $\Rvec_i$ are random positions.

It is worth noticing that in this scheme the LDOS $N(\Rvec,\omega)$ is calculated
for a {\it fixed} configuration of disorder (i.e. of impurities) and no
average over these configurations is taken. 
Furthermore we neglect the contributions from impurities that are
outside the field of view.  This results in Fourier-transformed
scattering potentials $\epsilon(\qvec)$, which could be sizably (and randomly)
momentum-dependent. Only owing to the rather large size of the $L\times L$
field of view and to the self-averaging character of this disordered system,
the $\epsilon(\qvec)$ functions turn out to be sufficiently smooth to preserve the
momentum structure of the $\Im (\Lambda_- (\qvec,\omega))$ encoding the interference effects
of the scattered QP's. In particular for an infinite system one would obtain
$|\epsilon(\qvec)|^2/N\approx \langle |\epsilon(\qvec)|^2\rangle/N=n_iV_0^2$
(where the angular brackets denote disorder average)\cite{capriotti}.
One could also calculate the LDOS power spectrum
by Fourier transforming the LDOS correlation function 
$\langle N (\Rvec,\omega)  N (0) \rangle $. Again, this procedure
would only allow to extract informations on the momentum structure
of $\Lambda(\qvec)$ if carried out on sufficiently large
grids such that the resulting $\epsilon(\qvec)$ is smooth.

We consider a system with a bare tight-binding band
$\epsilon_k=-2t(\cos k_x+\cos k_y)-4 t'\cos k_x\cos k_y -\mu_0$
($t$ and $t'$ are the nearest-neighbor and the next-nearest-neighbor
hopping parameters on a square lattice and $\mu_0$ is the
chemical potential) and a
$d$-wave superconducting gap 
$\Delta_k=\Delta_{0}(\cos k_x-\cos k_y)/2$.
In this way the QP dispersion is $E_\kvec=\sqrt{\epsilon_\kvec^2+\Delta_\kvec^2}$.
To compare our results with the experiments
of Refs. \onlinecite{hoffman1,hoffman2,mcelroy} 
we take the following parameter values, which, for $t\approx 150 \, meV$ are suitable for a
 ${\rm Bi_2Ca_2SrCu_2O_{8+x}}$ sample around optimal doping\cite{norman}:
$ t'=-0.3 t, \,\,\,
\Delta_{0} = 0.25 t =37.5 \, meV$.
For the sake of definiteness we take the doping $x=0.15$, for which 
we calculate the chemical potential to be $\mu_0 = -1.0 t$.

Fig. 1 displays the momentum dependence of the LDOS for this system and well
reproduces typical results of Ref. Capriotti {\it et al.}\cite{capriotti}.
To make the comparison with figures reported in experimental papers,
throughout this paper we will mark as darker the regions with larger
spectral intensity. Moreover, we orientate the momentum axes in such a way
that the Cu-O directions (usually taken as the $x=(1,0)$ and $y=(0,1)$
directions in theoretical papers) here are taken along the diagonals
of the figures. Notice also that, with the chosen orientation, 
the second Brillouin zone is visible in our figures. 

\begin{figure}
\includegraphics[angle=0,scale=0.85]{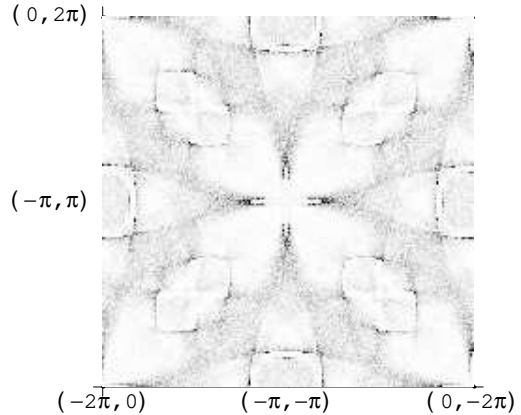}
\caption{Momentum-dependent LDOS for an homogeneous ${\rm Bi_2Ca_2SrCu_2O_{8+x}}$
superconducting system with $x=0.15$. The momenta are in units of the inverse 
lattice spacing.
To facilitate the comparison with experiments in this and subsequent figures
we rotate the Brillouin zone so that the Cu-O directions of the 
square $CuO_2$ planes are along the diagonals of the figure.
The tight-binding parameters are ($t=150\, meV$)
$ t'=-0.3 t; \,\, \mu_0 = -1.0 t \,; \,\,
\Delta_{0} = 0.25 t=37.5\, meV $, 
$\omega=-0.08t=-12\,meV$. The concentration of impurities
is $1\%$. 
}
\label{fig.1}
\end{figure}


The spectrum in Fig. 1 has been discussed in terms of the constant-energy curves 
of the quasiparticle dispersion relation $E_{\kvec}$.
Fixing the $E_{\kvec}$  identifies ``banana'' shaped  
curves in momentum space with large density of states at the 
extremities\cite{mcelroy,capriotti}.

These spectra 
are quite different in various respects from
the experimental ones. It is apparent that the
high-intensity regions are extended lines, which do not reproduce 
the spot-like shape of the experimental high-intensity regions.
One can show that large contributions to the spectra are obtained
when both denominators in Eq.~\ref{Lambda} are small.  
 Extended one-dimensional features are obtained when a translation
of $\qvec$ of one banana makes it tangent to another banana. 
Changing $\qvec$ keeping the two bananas tangent defines a 
 one-dimensional feature of high intensity. Crossing of two of these
 features produces a high intensity spot. However those spots are
 quite different from the experimental ones since they are clearly
 associated with the one-dimensional crossings.

Another important difference  is that there is no intensity around zero momentum. This lack
of spectral weight in the $\qvec =(0,0)$ region contrasts with the
presence of rather intense broad peaks appearing in the experimental
data.

Finally, the theoretical treatment considers a mesh coincident
with the atomic positions and correctly reproduces a spectrum, 
which is periodic in momentum space:  $N(\qvec,\omega)=
N(\qvec+{\bf G},\omega)$ with ${\bf G}$ a reciprocal lattice vector.
However, this feature is not  present in the experimental spectra,
where the peaks in the second Brillouin zone 
are suppressed with respect to their first  Brillouin zone partners. 
The presence of these form factors is
easily accounted for by the local space structure of the Wannier orbitals\cite{halperin}
and this effect is described in Appendix A. As shown in Fig. 2,
the introduction of Wannier orbitals does not significantly improve the
unrealistic appearance of the spectrum. This is most evident at low momenta,
where the non-point-like nature of the orbitals is obviously immaterial.
\begin{figure}
\includegraphics[angle=0,scale=0.85]{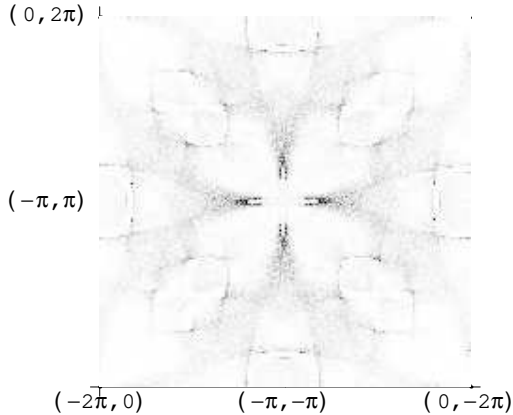}
\caption{Momentum-dependent LDOS spectra for a system with the
same parameters as in Fig. 1  and Gaussian Wannier orbitals with
width ${\sigma}=0.0625$ in units of square lattice spacing.
 The impurity concentration is
$1\%$ and the spectrum is taken at $\omega=-0.08t=-12\, meV$.}
\label{fig.2}
\end{figure}

In the next Sections, we will elaborate on the above expressions to consider
the inhomogeneous distribution of doping and the
consequent inhomogeneous distribution of the chemical
potential and of the superconducting gap.

\section{Effect of Mesoscopic inhomogeneities}

STM experiments
show large fluctuations in the gap amplitude over large 
length scales \cite{pan,lang}.
It is quite important to recognize that the size of these regions
is of several unit cells, $\xi \approx 13-15 A$, and is apparently 
unrelated with the average distance between impurities. 

Competition among different phases in strongly correlated systems
can give rise to mesoscopic inhomogeneities\cite{lor01I,lor01II,lor02}.
One can expect that this effect is enhanced by the inhomogeneous
distribution of the doping. Here we will not discuss the microscopic
mechanisms that can give rise to this effect but consider it as
granted and we analyze the consequence on the LDOS. 

Since the mesoscopic inhomogeneities involve several unit cells  
and differences can be substantial from one region to the other
 we choose an approach, which is analogous 
to the usual semiclassical
treatment of electrons in presence of slowly varying perturbations, 
where the electron distribution depends both
on space and momenta. In the same spirit 
we allow the Green functions to depend parametrically on 
the real-space region of the sample via the space dependence of both the
chemical potential $\mu=\mu(\Rvec)$ and the superconducting (SC) gap
$\Delta=\Delta(\Rvec)$. This gives rise to a local density of states
which depends explicitly on $\Rvec$ due to the conventional impurity
scattering, and implicitly trough the parametric dependence of the
gap and the chemical potential. 
\beqa
N(\qvec,\omega)&=&  \sum_{\Rvec} e^{-i\qvec\Rvec}N(\Delta(\Rvec),
\mu(\Rvec),\Rvec,\omega) \label{n0}
\eeqa
Here the sum is restricted to the $L\times L$ field of view  
and the first order expression of 
$N(\Delta(\Rvec),\mu(\Rvec),\qvec,\omega)$ is given by Eq.~\ref{sum}
but with the Green functions computed with the local value of  
$\mu(\Rvec)$ and $\Delta=\Delta(\Rvec)$. For example  ${\cal{G}}$
is given by Eq.(\ref{G11}), but with 
\begin{eqnarray}
\epsilon_k &=& -2t(\cos k_x+\cos k_y)-4 t'\cos k_x\cos k_y-\mu(\Rvec)\nonumber\\&&\\
\Delta_k &=& \Delta(\Rvec)(\cos k_x-\cos k_y)/2\nonumber
\end{eqnarray}
These expressions introduce a parametric dependence of ${\cal G}$ on the position
$\Rvec$ via the local values of the
chemical potential and of the maximum value of the $d$-wave SC gap $\Delta_0$
(however, to keep the notations simple, in the following we often do not explicitly 
indicate this $R$ dependence).
This approach allows for substantial differences in the gap and local
chemical potential from one region to the other and hence go beyond 
conventional perturbative formulations.

To realize a specific inhomogeneous distribution of
$\mu$  and  $\Delta$ we consider space fluctuations of the doping
around a given average value  $x$. Specifically we generate an 
inhomogeneous map characterized by doping fluctuations 
of 30\% (i.e. $x=x(\Rvec)$ locally ranges from $0.1$ to $0.2$) 
of typical size $\xi\sim 3-4$ lattice units.
We take this range of fluctuations as an estimate deduced
from the (larger) relative fluctuations of the gap observed
in the ${\rm Bi_2Ca_2SrCu_2O_{8+x}}$ sample of Ref. \onlinecite{pan}
and under the assumption that the gap and the doping are linearly
related.  For simplicity the smooth map was replaced by 
a mesa like function by determining contour levels of the 
smooth map and assigning to all points between two successive
contour levels a constant doping equal to the average value of the two
limiting contours. In this way the doping interval was (arbitrary)   
coarse-grained in five slices  and for each of the
five possible doping values the corresponding values of
the chemical potential and of the SC gap where calculated. 
In particular, starting from the given tight-binding structure,
the chemical potential was determined according to the local doping $x(\Rvec)$,
while for simplicity the maximum value of the $d$-wave gap was determined
by a linear rescaling with doping.
The resulting coarse-grained map is shown in Fig. 3 for a $200\times
200$ field of view.
 A direct comparison shows a close resemblance between our space
inhomogeneity map and the similar experimental figures of Refs. \onlinecite{pan,lang}.

\subsection{Zeroth order}
Even in the absence of impurity scattering, the inhomogeneities on the 
chemical potential and on the SC gap affect the LDOS
spectra. In particular if $\xi$ is the typical size of the domains
in which $\mu$ and $\Delta$ can  be taken as nearly constant,
it is quite natural to expect that a broadening 
of the spectra is obtained around the $q= (0,0)$ wave-vector, over a k-space
range of the order of $\xi^{-1}$. 
Specifically,
\beqa
N^{(0)}(\qvec,\omega)&=&  \sum_{\Rvec} e^{-i\qvec\Rvec}N^{(0)}
(\Delta(\Rvec),\mu(\Rvec),\omega) \label{n0b}
\eeqa
$N^{(0)}(\Delta(\Rvec),\mu(\Rvec),\omega)$ 
is the LDOS without impurities for a (homogeneous) system with 
SC gap $\Delta(\Rvec)$ and chemical potential $\mu(\Rvec)$:
$$
N^{(0)}(\Delta(\Rvec),\mu(\Rvec))=-\frac1{\pi}\Im
\int{\frac{d^2q}{(2\pi)^2}  {\cal G}(\Delta(\Rvec),\mu(\Rvec),\qvec,\omega)}
$$
where the Green function is taken in the absence of impurity scattering and is given by
the same expression of  Eq.(\ref{G11}), but with space-dependent parameters
$\mu(\Rvec)$ and $\Delta(\Rvec)$.

The numerical evaluation of $N^{(0)}(\qvec, \omega)$ [Eq. (\ref{n0b})] for the
specific distribution of inhomogeneities shown in Fig. 3 (upper panel), indeed produces a
rather strong peak around zero momentum (Fig. 3, lower panel).
This effect can be simply understood as follows.
The inhomogeneities in $\Delta(\Rvec)$  and $\mu(\Rvec)$ reflect in the DOS, 
which deviates from its
average value $N_0$. For illustrative purpose, we consider
an approximate linear dependence of the DOS from the energy in the
nodal approximation, so that, for small deviations of $\Delta$ and $\mu$ we find
\beq
N(\Delta(\Rvec),\mu(\Rvec))= N_0(\omega) \left[ 1+ \frac{\Delta(\Rvec)-\Delta_0}{\Delta_0} 
+\frac{\mu(\Rvec)-\mu_0}{\mu_0}+ ...\right]
\eeq
Here $\Delta_0\equiv \langle\Delta (\Rvec) \rangle$ and $\mu_0\equiv \langle\mu(\Rvec) 
\rangle$
are the average values of $\Delta$ and $\mu$ respectively.
We also assume a Gaussian distribution
\beq
P_X(\Rvec)={\cal N}e^{-\left[\xi^2\vert \nabla X(\Rvec) \vert^2 +X(\Rvec)^2\right]/A}  
\eeq
where $A$ is suitably chosen to give the expected size of the fluctuations of about 
thirty per cent.
Here $X(\Rvec)=\delta \Delta(\Rvec),\,\delta\mu(\Rvec)$, with 
$\delta \Delta(\Rvec)\equiv \Delta(\Rvec)-\Delta_0$ and $\delta\mu(\Rvec)
\equiv \mu(\Rvec)-\mu_0$.
The LDOS correlation function then takes the form
\beqa 
&&\langle N (\Delta(\Rvec),\mu(\Rvec))  N (\Delta(0),\mu(0)) \rangle =  \\
&N_0^2(\omega)& 
\left[ 1+ \frac{\langle \delta \Delta(\Rvec)\delta 
\Delta(0)\rangle }{\Delta_0^2} 
+\frac{\langle\delta\mu(\Rvec)\delta\mu(0)\rangle}{\mu_0^2}+ ...\right], \nonumber
\eeqa
Taking the Fourier transform, one obtains the LDOS power spectrum
\beq
P (\qvec,\omega) = 
N_0(\omega) \left[ \frac{A}{ \xi^2\vert \qvec  \vert^2 +1}
+\frac{A}{ \xi^2\vert \qvec  \vert^2 +1}
...\right].
\eeq
Therefore the $\mu$ and $\Delta$ distributions generate the
peak around $\qvec=(0,0)$ in the figures reported in the 
next subsections. As expected, we find that the width of this
peak scales as $\xi^{-1}$ showing that
this simple effect explains the central bump, which is not reproduced in 
the usual ``homogeneous'' approaches.
\begin{figure}
\includegraphics[angle=0,scale=0.85]{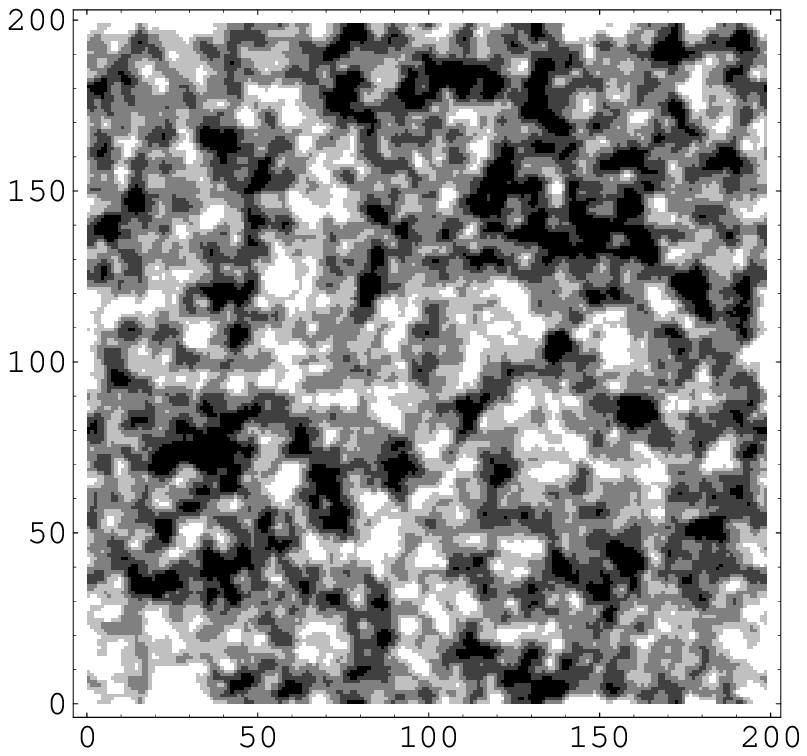}
\includegraphics[angle=0,scale=0.85]{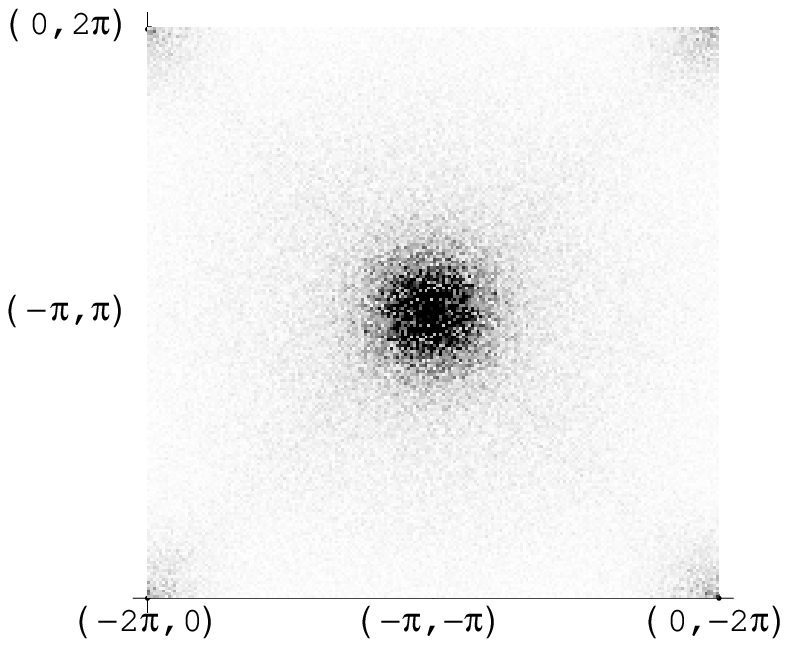}
\caption{Upper panel: Real-space map of a $200\times 200$ field of view with 
randomly distributed regions of tipical size $\xi\approx 3.5$
(in units of lattice spacing)
and five different values of doping and
corresponding values of gap and chemical potential (see text).
Lower panel: Momentum-dependent LDOS {\it without} impurity scattering
as it arises from Eq. (\ref{n0b}) with the mesoscopic distribution
of Fig. 3 (upper panel).}
\label{fig.3}
\end{figure}


\subsection{First order}
Besides the above simple zero-th order
effect, the local inhomogeneities affect the 
first-order corrections induced by
the scattering of the QP on the random impurities.
Therefore, in the presence of the
mesoscopic inhomogeneities, we recalculate the first-order
corrections to the LDOS 
\begin{equation}
 N^{(1)}(\qvec,\omega)=-\frac{1}{\pi}
\sum_\Rvec e^{-i\qvec\cdot \Rvec}\,\Im\left[\int \frac{d^2q'}{(2\pi)^2} 
e^{i\qvec'\cdot\Rvec}\epsilon(\qvec') \Lambda_-(\qvec',\omega)\right]
\end{equation}
where $\Lambda_-(\qvec,\omega)$ has the same form as in Eq. (\ref{Lambda}),
but with $R$-dependent normal and anomalous Green functions
according to their parametric dependence on $\Delta(\Rvec)$ and $\mu(\Rvec)$.
The sum over $\Rvec$ is to be intended over the
points of the $L\times L$ grid of the field of view.
We also consider the possibility of QP's scattering on 
magnetic impurities, for which the matrix representation is
given by
\begin{eqnarray*}
\hat{\epsilon}_m(\qvec)=\epsilon_m(\qvec)\left(
\begin{array}{cc}
1 & \phantom{-}0\\
0 & \phantom{-}1\\
\end{array}
\right)
\end{eqnarray*}
In the case of magnetic impurity scattering, the 
expression of $\Lambda$ for the LDOS of a spin-up electron reads
\begin{eqnarray}
&&{\Lambda_+}(\qvec,\omega)= \int \frac{d^2k}{(2\pi)^2}\nonumber\\ 
&&{\cal{G}}_R(\kvec,\omega)
{\cal{G}}_R(\kvec+\qvec,\omega)+
{\cal{F}}_R(\kvec,\omega){\cal{F}}_R(\kvec+\qvec,\omega)\nonumber\\
&&=\int \frac{d^2k}{(2\pi)^2} \frac{(\omega+\epsilon_\kvec)(\omega+\epsilon_{\kvec+\qvec})
+\Delta_\kvec \Delta_{\kvec+\qvec}}{(\omega^2-E_\kvec^2)(\omega^2-E_{\kvec+\qvec}^2)},
\label{lambdam}
\end{eqnarray}
Since an overall minus sign is obtained for spin-down electrons,
no contribution is obtained at first order for the total LDOS in the case of magnetic
impurities.

\subsection{Second order}
At second order in the impurity concentration, if we consider only 
scattering processes that occur on the same site, we have
\begin{eqnarray}
\label{nq2}
&&N^{(2)}(\qvec,\omega)=-\frac{1}{\pi}\sum_R e^{-i\qvec \Rvec}\times \\
\nonumber && \Im\left\{\int \frac{d^2p}{(2\pi)^2} e^{i\pvec\Rvec}\left[\Sigma_1(\pvec,\omega)
 \Lambda_-(\pvec,\omega)+\Sigma_2(\pvec,\omega) {\Lambda_+}(\pvec,\omega)\right]\right\}
\end{eqnarray}
where 
\begin{eqnarray}
\Sigma_1(\pvec,\omega)&=&\left(\epsilon^2(\pvec)+\epsilon_m^2(\pvec)\right)
\sum_\kvec \frac{\epsilon_\kvec}{\omega^2-E_\kvec^2}\label{sigma1}\\
\Sigma_2(\pvec,\omega)&=&\left(\epsilon^2(\pvec)+\epsilon_m^2(\pvec)\right)\sum_\kvec 
\frac{\omega}{\omega^2-E_\kvec^2}\label{sigma2}
\end{eqnarray}
and $\Lambda_\pm$ are given by Eq.(\ref{Lambda}) and 
Eq.(\ref{lambdam}) respectively.

Various observations are in order here. First of all
it is found that at second order, a finite contribution to the 
LDOS in momentum space is obtained also in the case of magnetic scattering.
Notice also that, contrary to the first-order case, the real parts
of $\Lambda_-$ and ${\Lambda_+}$ both contribute to the LDOS.
Finally  we notice from Eqs. (\ref{nq2})-(\ref{sigma2}) that second-order
QP interference processes may contribute to the LDOS at zero momentum
$N({\bf q}=(0,0))$.
This contribution adds to the zeroth-order peak arising from the 
mesoscopic inhomogeneities described in Sec. III.B and may be present
even in the absence of such inhomogeneities. This can be seen by
taking $\mu(\Rvec)=\mu_0$ and $\Delta(\Rvec)=\Delta_0$. 

Although at $\qvec =(0,0)$,  the first-order contribution
to $\Lambda_-(0)$ is small [see Fig. 4 (upper panel) and 
Appendix B], $\Lambda_+(0)$ 
does not vanish and contributes to $N^{(2)}(0)$ (both in the absence and in the 
presence of magnetic scattering). Therefore second-order scattering processes
could contribute to the intense (rather
broad) peak experimentally obtained for $N(\qvec,\omega)$ at ${\bf q}=(0,0)$.
However, we checked from the relative intensity of the peaks at finite momenta
that the scattering processes are weak and the second-order processes are
not strong enough to explain the rather large intensity of the
peak at $\qvec =(0,0)$. This strongly indicates that the 
zero-order contribution from the mesoscopic inhomogeneities is the
main source of the large intensity of the zero-momentum peaks
in the experiments.

\section{Microscopic Spatial charge modulations}
The observation (see below) that calculations 
of the STM spectra in terms of interfering scattered
quasiparticles do not account for the observed weight of
some specific spectral features
motivated a further enrichment of our treatment. Specifically
we considered a physically different effect
arising from microscopic charge modulations. Since long time
it has been suggested that the superconducting cuprates
might be close to an instability leading to the  spatial
ordering\cite{referenzestripes} likely in the form of fluctuating
onedimensional charge textures (the so-called stripes), which 
can even acquire a slow critical dynamics \cite{CDG} around
optimal doping and above. The scale of these textures is of a few
lattice spacing thus we call them ``microscopic'' as opposed to the
mesoscopic inhomogeneities consider above. 
The proximity to the microscopic instability
can naturally reflect itself in a large enhancement of the 
charge susceptibility at specific wave-vectors ${\bf Q}_{ch}$
(obviously corresponding to the charge order) and to related structures
in the momentum-dependent dielectric function. In particular,
it is natural that the external potential introduced by
static impurities is substantially screened by the nearly
unstable electron liquid and acquires some structure in 
its momentum dependence. One can see that close to the charge
instability, the static charge susceptibility acquires a nearly polar 
form\cite{notaperiodicity}
\beq 
\chi({\bf q}) = \frac{\chi_0 ({\bf q})}{1-V({\bf q})\chi_0({\bf q})}
\propto \frac{\chi_0 ({\bf q})}{\vert {\bf q}-{\bf Q}_{ch} \vert^2
+ \xi_{ch}^{-2}}
\eeq
where $\xi_{ch}$ is the correlation length for charge fluctuations.
Once this charge susceptibility is introduced in the dielectric constant,
one finds that the impurity potential is screened as
\beq {\tilde V}_0({\bf q})= V_0 \left(1 + \sum_{Q_{ch}} 
\frac{1}{(\vert \qvec-{\bf Q}_{ch}\vert^2+\xi^{-2}}\right).
\label{v0q}
\eeq
For simplicity, we assume here $V_0\chi_0(\qvec={\bf Q}_{ch})\sim 1$
and we take $\xi_{ch}^{-1}=0.25$ inverse lattice spacing.
This screening gives rise to a strongly momentum-dependent impurity potential.
Quite obviously, once this screened impurity potential is inserted in the
expressions for $N(\qvec,\omega)$, it will filter the momentum dependence of
the LDOS emphasizing the intensity at $\qvec \approx {\bf Q}_{ch}$.

\section{Results}
We here describe the
effects in the calculated  momentum-dependent LDOS arising from
the progressive introduction of the mesoscopic inhomogeneities 
and of the spatial charge modulations according to the
scheme described in Section III and IV. To compare with the 
results reported in Figs. 1 and 2 for an homogeneous system,
we keep considering the tight-binding model 
with the same parameters.
Again the value of $\mu_0$ corresponds to a doping $x=0.15$.
Besides the previously considered effect of 
the local Wannier structure of the orbitals, we introduce 
a random inhomogeneous (mesoscopic) doping distribution 
according to the scheme of Subsection III.A.
Fig. 4 represents a momentum-space LDOS including up to second-order scattering.
As it can be clearly seen, the space inhomogeneity is quite effective in modifying the
STM spectra (a) by introducing substantial spectral weight around zero momenta, 
and (b) by broadening the curve-like spectral features. In particular, these latters
acquire a ``fuzzier'' appearance, which emphasizes the regions of stronger intensity
thereby producing more spot-like features and rendering the calculated spectra closer 
to the experimental data. We also notice that this effect arises from the disorder
in gap and chemical potential and is therefore physically quite different from
(and its effects quite more pronounced than)  the
impurity disorder considered in the appendix of Ref. \onlinecite{capriotti}.

\begin{figure}
\includegraphics[angle=0,scale=0.85]{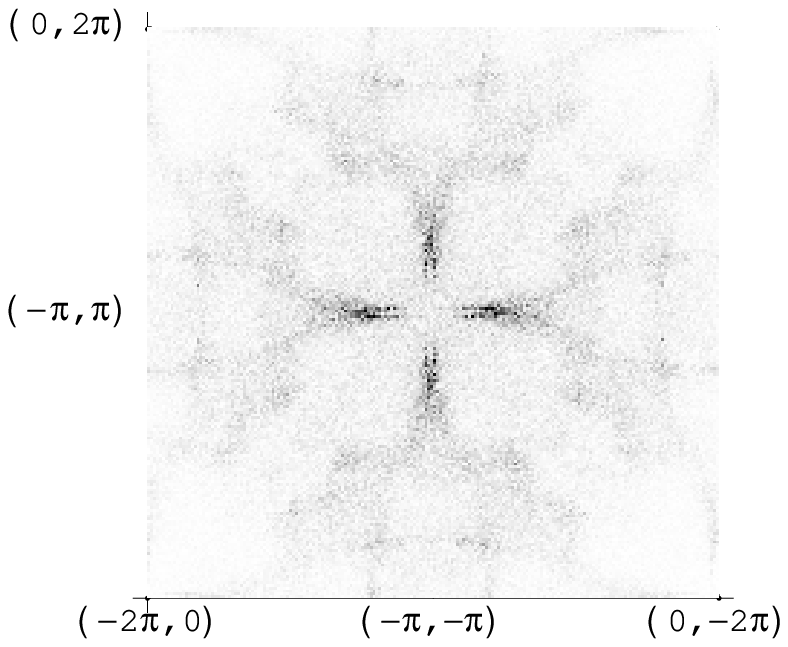}
\includegraphics[angle=0,scale=0.85]{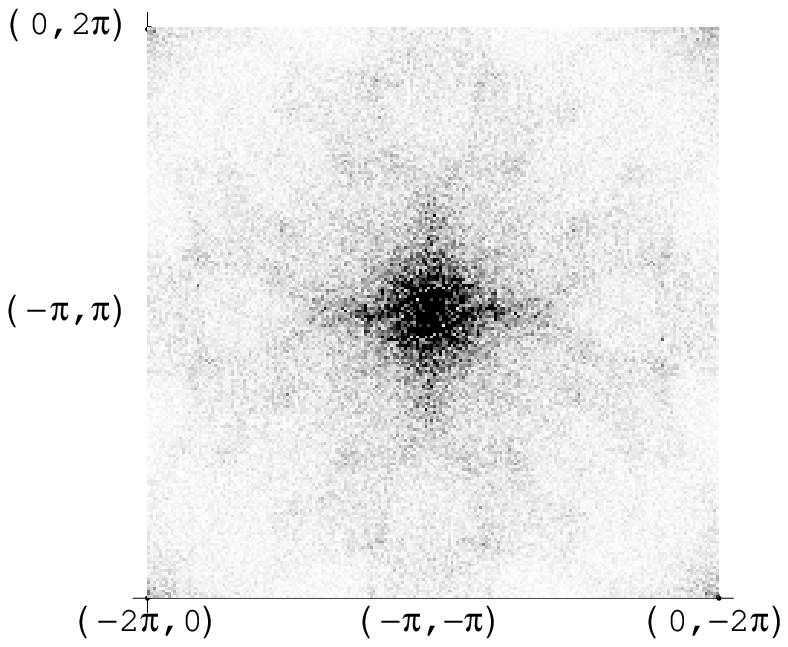}
\caption{Upper panel: First-order-only contribution to the
momentum-dependent LDOS spectra for a system with the
same parameters as in Fig. 1 and 2 (Wannier orbitals are also considered).
The chemical potential and the SC gap
are randomly distributed according to the real-space map of Fig. 3 (upper panel).
 The impurity concentration is
$1\%$ and the spectrum is taken at $\omega=-0.08t=-12\, meV$.
Lower panel: Momentum-dependent LDOS spectra with the impurity scattering calculated
including the zeroth, first and second order with $\epsilon=0.8t$. All other parameters are the
same as in the upper panel, }
\label{fig.4}
\end{figure}
Despite this substantial improvement, a closer comparison with the data of 
Refs. \onlinecite{hoffman2} and \onlinecite{mcelroy} (cf. Fig. 5 (lower panel) 
shows that the calculations of Fig. 4 fail in 
reproducing some of the rather intense features experimentally detected.
Specifically, strong spectral features are observed along the Cu-O-Cu
directions (i.e. along the diagonals of Figs. 4 and 5, see the experimental 
intensities schematically reported in Fig. 5, lower panel)
at wave-vectors corresponding to a four-unit-cell modulation
$\vert{\bf q} \vert= 0.25 \times (2\pi /a_0)$ ($a_0$ is the lattice unit in the
supposedly square $CuO_2$ planes) \cite{howald0,howald2,mcelroy}, which are
to weak in Fig. 4 (lower panel). Therefore, following the scheme of Section IV,
we consider the effects of screening on the impurity potential. According
to the suggestion that the cuprates are close to a charge instability, this
screening is taken to be strongly momentum dependent, as a result of the
strong charge susceptibility of the electron liquid at momenta 
$\qvec \approx {\bf Q}_{ch}$. The resulting calculated spectrum is reported in
Fig. 5 (upper panel), where the too weak peaks in the $(\pm 1,0),(0,\pm 1)$
directions (along the diagonals of the figures)
 are now strongly enhanced. The comparison with experimental intensities
is more direct in Fig. 5 (lower panel), where we use dots to depict the regions
where experimental spectra display the strongest features. It is clear that
the effect of momentum-dependent screening on the impurity potential is substantial
and important to correctly reproduce {\it all} the experimental structures.

\begin{figure}
\includegraphics[angle=0,scale=0.85]{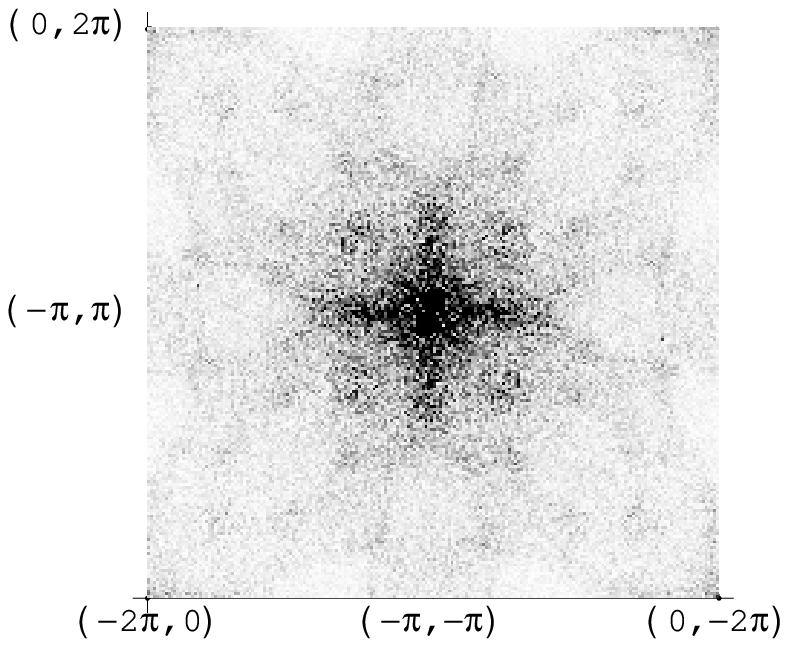}
\includegraphics[angle=0,scale=0.85]{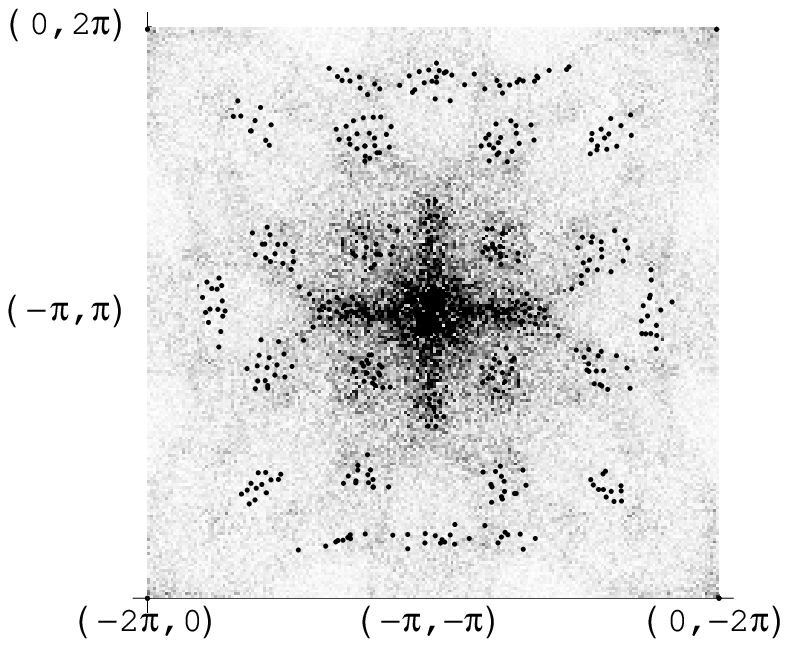}
\caption{Upper panel: Momentum-dependent LDOS spectrum including the effects of 
mesoscopic inhomogeneities and the impurity scattering
up to second order with $\epsilon=0.8t$.
We also considered a momentum-dependent screened
impurity potential according to Eqs. (\ref{v0q}) and .
(\ref{v0q}) with $\xi_{ch}^{-1}=0.25$ in units of inverse lattice spacing.
The impurity concentration is
$1\%$ and the spectrum is taken at $\omega=-0.08t=-12\, meV$.
Lower panel: same as in the upper panel, but with the addition of
dots  marking intense features in the 
experiments of Ref. \onlinecite{mcelroy}}
\label{fig.5}
\end{figure}

\section{Conclusions}

The results displayed in the previous sections allow to draw some
general conclusions concerning the mechanisms leading to the
formation of (point-like) structures in the LDOS as obtained
in Fourier-transformed STM spectra. 

First of all, disregarding the zero-order inhomogenity effects
acting at low momenta, we considered the intensity of the finite-momentum
peaks. From the comparison of the first and second-order
contributions with experimental spectra,
we can draw the conclusion that the
QP scattering due to impurities is rather weak. Therefore
our perturbative approach is appropriate and the first-order
processes already account well for the part of the
spectra, which can be attributed to the QP interference.
When this mechanism at first order fails in reproducing the
all relevant features of the spectra, the
second-order processes do not substantially improve the 
calculations. The same holds true as far as the inclusion
of magnetic impurities is concerned.
We also took in consideration the extended
character of the Wannier orbitals, which partially modifies the 
spectra at large wave-vectors accounting for the 
weakening of the spectral features
at large momenta. Nevertheless the calculated spectra
preserve their unrealistic features.

A substantial improvement in the
theoretical calculation is represented by the inclusion
of space inhomogeneities both in the SC gap and in the 
chemical potential as arising from doping inhomogeneities.
As far as the impurity scattering is concerned, the main effect
of inhomogeneities is to blur the curve-shaped spectral features
promoting a more point-like appearance of the high intensity
regions of the spectra. This is quite easily interpreted in terms
of the (by now) standard arguments related to the relevance of
some specific wave-vectors joining the parts of the electronic
(d-wave gapped) spectra at fixed energy bias (see, e.g., Fig. 1
of Ref. \onlinecite{mcelroy}). The  contours of constant-energy 
in the quasiparticle spectra have the well-known ``banana'' shape
and in the absence of disorder the momentum space regions with large
intensity are extended curves. However, by introducing local gap
fluctuations, the length of the ``bananas'' fluctuates, while the transversal,
i.e. the width, fluctuations of the ``bananas'' are instead less
pronounced due to the very elongated shape of the constant-energy contours.
In this case one sees that the gap fluctuations produce already a 
marked blurring of the LDOS peaks. The introduction of the local
chemical potential fluctuations, instead, tends to produce shifts
in the Fermi surface and in the quasiparticle spectra, which are
most pronounced along the Fermi velocities $v_F$. This locally
induces transversal shifts of the ``bananas'', further blurring
the LDOS spectral features. As a result, the unrealistic curve-like
regions of high LDOS intensity in momentum space are transformed
in spot-like features, which have a closer appearance to the 
experimental spectra. This is the first main outcome of our work.
Despite this success, however, there are some specific spectral 
features at some specific momenta, 
$\vert{\bf q} \vert= 0.25 \times (2\pi /a_0)$ (close to the
so-called $\qvec_1$ in the ``octet'' model of Ref. \onlinecite{mcelroy})
which we did not reproduce with the proper 
intensity within the QP interpherence mechanism. 
It was already observed in Ref. \onlinecite{howald2}, that theoretical
calculations based on interference effects between impurity-scattered
quasiparticles underestimate the above spectral features. 
Moreover, recent experiments \cite{yazdani}
 in the pseudogap region of ${\rm Bi_2Ca_2SrCu_2O_{8+x}}$ 
above have shown the persistence of LDOS peaks at some momenta, which 
can hardly be attributed to interphering QP's and could rather be
attributed to some spatial order in the charge and/or in the spin channels.
Our findings strengthen this observation showing that the several improvements
at large (Wannier functions) and small (zero-order effects) wave-vectors, as well as
second-order calculations do not change this conclusion.
Therefore,  aiming to reproduce {\it all} the features of the STM spectra, 
we included a momentum-dependent screening due to a (supposedly strong)
charge susceptibility at some specific wave-vectors. As described in Fig. 5
above, the close resemblance of the calculated spectra to the experiments suggests
that a strong tendency of the cuprates to order spatially at some specific
wave-vectors  might well be present in the real systems and coexist with
$d$-wave quasiparticles.
This important indications represents the second main outcome of our work.

After this work was completed, we became aware of more recent STM experiments\cite{mcelroy2}
where two different spatially separated regions were identified in underdoped
${\rm Bi_2Ca_2SrCu_2O_{8+x}}$ samples. In (more metallic) regions with
large coherence peaks (which are the large
majority around optimal doping) the dispersive feature related to
the $\qvec_1$ vector of the ``octet'' model has a smooth
intensity variation. 
 On the other hand, in (less metallic) regions
where the coherence peaks are absent, the dispersive spectral feature
related to  $\qvec_1$ acquires an additional sizable intensity, when
 $\qvec_1$ is close to the incommensurate
charge ordering momenta ${\bf q}\approx 0.25 \times (2\pi /a_0)(\pm 1,0)$ and
${\bf q}\approx 0.25 \times (2\pi /a_0)(0,\pm 1)$. This 
suggest a substantial charge ordering in these regions. From this point of view
our Fig. 5 represents a ``superposition'' of 
the spectra from these spatially separated regions.

\acknowledgments
We  thank L. Capriotti and C. Di Castro for useful suggestions and discussions.
We acknowledge financial support from the MIUR Cofin 2003 prot. $2003020239_006$

\appendix
\section{Local density of states with Wannier functions}
The first striking difference with experiments is that $\Lambda_-(\qvec,\omega)$ is periodic
in the reciprocal space, whereas the experimental spectra are not. 
One first natural step toward a more realistic description of the STM
spectra can be carried out by considering the real space structure
of the Wannier orbitals. In this way the short-distance 
description around each lattice point is improved, leading to 
a refinement of the calculated STM spectra at large wave-vectors.
Defining 
$\phi_\Rvec(\rvec)$ a Wannier function with center in $R$, a vector of the lattice,
the electron Green functions can be expanded on this basis so that
the LDOS at the first order in the impurity concentration can be written in the most general way
\begin{eqnarray}
\hspace{-0.4cm} N(\rvec,\omega)&=&\Im\int d\rvec_0 \,G(\rvec,\rvec_0)\,\epsilon(\rvec_0)
\,G(\rvec_0,\rvec,\omega)\\
\nonumber&=&\Im
\sum_{\kvec_1,\kvec_2}\sum_{\Rvec_1,\Rvec_2}\sum_{\Rvec'_1,\Rvec'_2} 
\left\{ G(\kvec_1,\omega))\,G(\kvec_2,\omega)) \times \right.\\
\nonumber && e^{i\kvec_1\cdot(\Rvec_1-\Rvec'_1)}
\,e^{i\kvec_2\cdot(\Rvec_2-\Rvec'_2)}\phi_{\Rvec_1}(\rvec,\omega)\, 
\phi^*_{\Rvec'_2}(\rvec)\times \phantom{\int}\\ 
\nonumber &&
\left.\int d\rvec_0\, \phi^*_{\Rvec'_1}(\rvec_0)\,\epsilon(\rvec_0)\,
\phi_{\Rvec_2}(\rvec_0)\right\}
\end{eqnarray}
where $GG={\cal{G}}{\cal{G}}$ in the normal phase and 
$GG={\cal{G}}{\cal{G}}-{\cal{F}}{\cal{F}}$ in the superconducting phase
If one assumes that the impurity potential  $\epsilon(\rvec)$ is only defined 
on the lattice sites $R$, then
\begin{eqnarray}
\int d\rvec\, \phi^*_{\Rvec'_1}(\rvec)\,\epsilon(\rvec)\,\phi_{\Rvec_2}(\rvec)\simeq 
\delta_{\Rvec'_1,\Rvec_2}\sum_{\qvec} e^{-i\qvec\cdot \Rvec_2}\, \epsilon(\qvec) 
\end{eqnarray}
and one obtains the following expression for the LDOS
\begin{eqnarray}
N(\rvec,\omega)&=&\Im \sum_{\kvec,\qvec}\sum_{\Rvec_1,\Rvec'_2}G(\kvec,\omega)
\,G(\kvec+\qvec,\omega)\,e^{i\kvec\cdot \Rvec_1} \\
\nonumber &&e^{-i(\kvec+\qvec)\cdot \Rvec'_2}\,\epsilon(\qvec)\, 
\phi_{\Rvec_1}(\rvec) \,\phi^*_{\Rvec'_2}(\rvec)
\end{eqnarray}
If one further assumes that the Wannier orbitals are rather localized
around each lattice site, the following approximation becomes applicable
\begin{eqnarray}
  \phi_{\Rvec_1}(\rvec)\,\phi^*_{\Rvec'_2}(\rvec)\simeq \delta_{\Rvec_1,\Rvec'_2}|
\phi_{\Rvec_1}(\rvec)|^2 
\end{eqnarray}
and taking advantage of the property of the Wannier functions
\begin{eqnarray}
\phi_{\Rvec}(\rvec)=\phi_{\Rvec+\Rvec'}(\rvec+\Rvec'),
\end{eqnarray}
the Fourier transformation of $N(\rvec,\omega)$,
\begin{eqnarray}
N(\qvec,\omega)=\int d\rvec\, e^{-i \qvec\cdot \rvec} N(\rvec,\omega)
\end{eqnarray}
can be written as
\begin{eqnarray}
N(\qvec,\omega)= F(\qvec) \sum_\Rvec 
e^{-i \qvec\cdot \Rvec}\,\Im  \sum_{\qvec}e^{i \qvec\cdot \Rvec} 
\, \epsilon(\qvec) \,\Lambda_-(\qvec,\omega)
\end{eqnarray}
Here the form factor 
\begin{eqnarray}
F(\qvec)= \int d\rvec e^{-i \qvec\cdot \rvec}\,|\phi_{0}(\rvec)|^2 
\simeq e^{-(q_x^2+q_y^2)\,\sigma}
\label{F}
\end{eqnarray}
has been made explicit by assuming a Gaussian shape of the (rather localized)
Wannier orbital, with  $\sigma$ proportional to the spacial variance of the
Wannier functions. 

It is quite simple to check that,
if Wannier functions are delta-like, $\sigma\rightarrow 0$ and $F(\qvec)\rightarrow 1$,
and the usual expressions of Ref. \onlinecite{capriotti} are recovered.

\section{Local density of states in the point $(0,0)$}
We numerically find [Fig. 4 (upper panel)] that 
$\Lambda_-(\qvec=0,\omega)$ is quite small. This numerical finding 
can acquire an analytic support, from a calculation at low $\omega$,
where a nodal approximation is justified. In this case we can show that
$\Lambda_-(\qvec=0,\omega)$ vanishes. Let us consider 
$$\Lambda_\pm (\qvec,\omega)=
\int \frac{d{\kvec }}{(2\pi)^2} \frac{(\omega+\epsilon_\kvec)
(\omega+\epsilon_{\kvec+\qvec})\mp 
  \Delta_\kvec\Delta_{\kvec+\qvec}}{(\omega^2-E^2_\kvec)(\omega^2-E^2_{\kvec+\qvec})}$$
At first order only $\Lambda_-$ enters the calculation for non-magnetic
impurities, while only ${\Lambda_+}$ is to be used for magnetic impurities.
In this latter case the summation over spins leads to a cancelation
and no contribution is obtained at first order.
On the other hand, at second order, both $\Lambda_-$ and ${\Lambda_+}$ 
are needed both for magnetic and non-magnetic impurities [cf. Eqs. (\ref{nq2})-(\ref{sigma2}].
In order to calculate the local density of states at  $\qvec=(0,0)$
both at first and second order
we consider
\begin{eqnarray}
\Lambda_\pm ({\bf 0},\omega)=\int 
\frac{d{\kvec}}{(2\pi)^2}\frac{(\omega+\epsilon_\kvec)^2\mp 
\Delta_\kvec^2}{(\omega^2-E^2_\kvec)^2} 
\label{lamb0}
\end{eqnarray}
If we change the variables of integration near a node 
$$\int \frac{d{\kvec}}{(2\pi)^2}\longrightarrow \int_0^{2\pi}  
\frac{d\theta}{(2\pi)} \int \frac{d\rho\, \rho}{2\pi v_F v_\Delta}$$
so that
\begin{eqnarray*}
&&\epsilon_\kvec=\rho\, \cos\theta\\
&&\Delta_\kvec=\rho\, \sin\theta,
\end{eqnarray*}
the Eq. (\ref{lamb0}) becomes
\begin{eqnarray}
\Lambda_\pm (\qvec,\omega)&=&4\int_0^{2\pi}  \frac{d\theta}{(2\pi)} \times \nonumber \\
&&\int \frac{d\rho^2}{4\pi v_F v_\Delta}
 \frac{\omega^2+\rho^2\left((\cos\theta)^2\mp (\sin\theta)^2\right)}{(\omega^2-\rho^2)^2}.
\end{eqnarray}
For $\Lambda_-({\bf 0})$ we have
\beqa
\Lambda_-({\bf 0})&=&\int_0^{2\pi}\frac{d\theta}{(2\pi)} \int \frac{d\rho^2}{\pi v_F v_\Delta} 
\frac{\omega^2+\rho^2 \cos2\theta}{(\omega^2-\rho^2)^2}\nonumber \\
&=& \frac{1}{\pi v_F v_\Delta}
\int d\rho^2\frac{\omega^2}{(\omega^2-\rho^2)^2}
\eeqa
For ${\Lambda_+}({\bf 0})$ we have instead
$${\Lambda_+}({\bf 0})=\frac{1}{\pi v_F v_\Delta}\int d\rho^2 
\frac{\omega^2+\rho^2}{(\omega^2-\rho^2)^2}$$
The local density of states in the long wavelength limit in $\qvec=(0,0)$
 and in the presence of non magnetic and magnetic impurity is given by 
\begin{eqnarray}
&& \frac{1}{\pi v_F v_\Delta}\, \Im \int dz \frac{\omega^2}{(\omega^2-z)^2}=0\\
&& \frac{1}{\pi v_F v_\Delta}\,\Im \int dz \frac{\omega^2+z}{(\omega^2-z)^2}=\frac{1}{v_F v_\Delta}
\end{eqnarray}
with a double pole in $\omega^2+i0^{+}$.
This means that the spatial average of $n(\Rvec)$ at first order is zero for a non magnetic impurity,
while is finite for a magnetic impurity {\it at fixed spin}. In this latter case it could
provide a direct measure at low energies of the product between Fermi and gap velocities.
However, as already stated at the end of Sect. III.C, summing over spins
one again obtains a vanishing contribution. On the other hand, at
second order both $\Lambda_-$ and ${\Lambda_+}$ contribute to the scattering
in the presence of magnetic and/or non-magnetic scattering.



\begin{thebibliography}{99}
\bibitem{notaperiodicity}To render the following expressions simpler we 
neglected the periodic character of the momentum dependence, but we explicitely considered
it in the actual numerical calculations.
\end{thebibliography}
\end{document}